\documentclass[11pt,a4paper]{article}
\usepackage{amsmath,amssymb}
\usepackage{latexsym}

\newtheorem{theorem}{Theorem}

\newtheorem{proposition}{Proposition}
\newtheorem{corollary}{Corollary}

\author{H{\aa}kan Andr\'{e}asson\\
Mathematical Sciences\\ University of Gothenburg\\
        Mathematical Sciences\\Chalmers University of Technology\\
        S-41296 G\"oteborg, Sweden\\
        email: hand@chalmers.se\\
        \ \\
        Christian G.~B\"ohmer\\
        Department of Mathematics\\ University College London\\
        Gower Street, London, WC1E 6BT, UK\\
        email: c.boehmer@ucl.ac.uk}

\title{Bounds on $M/R$ for static objects with a positive cosmological constant}

\date{}

\begin{document}

\maketitle

\begin{abstract}
We consider spherically symmetric static solutions of the Einstein equations with a positive cosmological constant $\Lambda,$ which are regular at the centre, and we investigate the influence of $\Lambda$ on the bound of $M/R,$ where $M$ is the ADM mass and $R$ is the area radius of the boundary of the static object. We find that for any solution which satisfies the energy condition $p+2p_{\perp}\leq\rho,$ where $p\geq 0$ and $p_{\perp}$ are the radial and tangential pressures respectively, and $\rho\geq 0$ is the energy density, and for which $0\leq \Lambda R^2\leq 1,$ the inequality
\begin{equation*}
  \frac{M}{R}\leq\frac29-\frac{\Lambda R^2}{3}+\frac29 \sqrt{1+3\Lambda R^2},
\end{equation*}
holds.
If $\Lambda=0$ it is known that infinitely thin shell solutions uniquely saturate the inequality, i.e. the inequality is sharp in that case. The situation is quite different if $\Lambda>0.$ Indeed, we show that infinitely thin shell solutions \textit{do not} generally saturate the inequality except in the two degenerate situations $\Lambda R^2=0$ and $\Lambda R^2=1$. In the latter situation there is also a constant density solution, where the exterior spacetime is the Nariai solution, which saturates the inequality, hence, the saturating solution is non-unique. In this case the cosmological horizon and the black hole horizon coincide. This is analogous to the charged situation where there is numerical evidence that uniqueness of the saturating solution is lost when the inner and outer horizons of the Reissner-Nordstr\"{o}m solution coincide.
\end{abstract}

\section{Introduction}

A fundamental question concerning spherically symmetric relativistic static objects is to determine an upper bound on the gravitational red shift. In the case with a vanishing cosmological constant this is equivalent to determining an upper bound on the compactness ratio $M/R,$ where $M$ is the ADM mass and $R$ the area radius of the boundary of the static object. Buchdahl's theorem \cite{Bu1} is well-known and shows that a spherically symmetric isotropic object for which the energy density is non-increasing outwards satisfies the bound
\begin{align}\label{Buchdahl}
\frac{M}{R}\leq\frac{4}{9}.
\end{align}
The inequality is sharp, but the solution which saturates the inequality within the class of solutions considered by Buchdahl violates the dominant energy condition and is therefore unphysical. Moreover, the assumptions that the pressure is isotropic, and the energy density is non-increasing, are quite restrictive. In \cite{An1} it was shown that the bound (\ref{Buchdahl}) holds generally, i.e. independently of the Buchdahl assumptions, for the class of solutions which satisfy the energy condition
\begin{align}\label{energycondition}
p+2p_{\perp}\leq \rho.
\end{align}
Here $p\geq 0$ is the radial pressure, $p_{\perp}$ the tangential pressure and $\rho\geq 0$ the energy density. It should be pointed out that (\ref{energycondition}) is natural and is e.g. satisfied for solutions of the Einstein-Vlasov system, cf. \cite{An4} for a review on the  Einstein-Vlasov system. Moreover, Bondi uses this condition in his study on anisotropic objects in \cite{Bo2}. It was in addition shown in \cite{An1} that (\ref{Buchdahl}) is sharp within this class of solutions and that the saturating solution is unique, it is an infinitely thin shell. Since an infinitely thin shell is singular this should be interpreted in the sense that $M/R\to 4/9$ for a sequence of regular shell solutions which approach an infinitely thin shell. That arbitrarily thin shell solutions do exist has been shown for the Einstein-Vlasov system in \cite{An2}, cf. also the numerical study \cite{AR2}. Since the saturating solution satisfies (\ref{energycondition}), it satisfies in particular the dominant energy condition. Note on the other hand that it neither satisfies the isotropy condition nor the assumption on the energy density in the Buchdahl assumptions. An alternative proof to the one in \cite{An1} was given in \cite{KS}. The advantage being that it is shorter and more flexible since it allows for other energy conditions than (\ref{energycondition}). The disadvantage is the proof of sharpness which does not show that the saturating solution is unique. Moreover, the saturating solution constructed in \cite{KS} has features which e.g. solutions of the Einstein-Vlasov system cannot have.

In the present study we investigate the influence on the bound of $M/R$ in the presence of a positive cosmological constant. A previous study in this case was carried out in \cite{BO1,BO2,Ba,BH2,BH} under the assumptions used by Buchdahl. Here we relax the Buchdahl assumptions and impose the condition (\ref{energycondition}), and we find that if $\Lambda R^2<1,$ the inequality
\begin{equation}\label{ineqlambda}
  \frac{M}{R}\leq\frac29-\frac{\Lambda R^2}{3}+\frac29 \sqrt{1+3\Lambda R^2},
\end{equation}
holds generally for this class of solutions. In fact, the inequality holds in the interior of the static object as well and not only at its boundary, cf. Theorem~\ref{theorem1}. The proof of this part is an adaption of the method in \cite{KS}. The natural question of sharpness of (\ref{ineqlambda}) is also addressed but we have not been able to give a completely satisfying answer. However, we give a detailed analysis of sharpness for the class of infinitely thin shell solutions. Since an infinitely thin shell solution saturates the Buchdahl inequality, i.e. the case $\Lambda=0,$ as well as the inequality derived in \cite{An5} for charged static objects, it is quite natural to investigate if infinitely thin shell solutions saturate (\ref{ineqlambda}) as well. We show, by using ideas from \cite{An3}, that generally this is not the case except in two situations. The most interesting exception being an infinitely thin shell solution for which $\Lambda R^2\to 1$. This case belongs to the boundary of the domain we consider. From (\ref{ineqlambda}) it follows that $M/R\to 1/3,$ as $\Lambda R^2\to 1,$ for such a shell since it saturates the inequality. This is considerably lower than $4/9$ in (\ref{Buchdahl}), but the presence of a cosmological constant changes the expression for the gravitational red shift which, as a matter of fact, becomes unbounded in this case.

It is interesting to compare the result obtained in the present paper by the result in \cite{BO2} where constant density solutions where considered. In the domain $0\leq \Lambda R^2\leq 1,$ the following inequality
\begin{equation}\label{rhoconst}
\frac{M}{R}\leq\frac29+\frac29\sqrt{1-\frac{3\Lambda R^2}{4}},
\end{equation}
is derived in \cite{BO2} for constant density solutions. The pressure is given by the Tolman-Oppenheimer-Volkov equation, and the condition (\ref{energycondition}) is not necessarily satisfied in this case. Let us first point out that the inequality (\ref{ineqlambda}) admits larger values of $M/R$ than the inequality (\ref{rhoconst}) when $0<\Lambda R^2<1.$ At the end points of the interval the two inequalities however agree, i.e. $M/R\leq 4/9$ when $\Lambda R^2=0$ and $M/R\leq 1/3$ when $\Lambda R^2=1.$ As mentioned above, in this work we construct an infinitely thin shell solution which saturates the inequality when $\Lambda R^2=1.$ In \cite{BO2} a sequence of isotropic constant density, $\rho_0$ say, perfect fluid spheres with increasing radius is considered, where the radius can be controlled by the density. It turns out that in this sequence there is exactly one solution which saturates the inequality when $\Lambda R^2=1,$ namely the situation where $\Lambda = 4\pi \rho_0$ and the exterior spacetime is the Nariai solution \cite{Na,BO1,BO2}, which satisfies the energy condition provided $3p_c \leq \rho_0$, where $p_c$ is the central pressure. Hence, the saturating solution is non-unique when $\Lambda R^2=1.$ In the case of a constant density solution with $\Lambda = 4\pi \rho_0,$ it is exactly when $\Lambda R^2=1$ that the cosmological horizon and the black hole horizon coincide. It is quite striking that a similar result holds in the case of charged solutions. In \cite{AER} numerical evidence is given that two classes of saturating solutions to the inequality derived in \cite{An5} for charged solutions exist. This happens exactly when the inner and the outer horizon of the Reissner-Nordstr\"{o}m black hole coincide.

The outline of the paper is as follows. In the next section we set up the system of equations and present our two main results. Section 3 and 4 are devoted to their proofs. 

\section{Set up and main results}

We consider the static and spherically symmetric line element in Gauss coordinates relative to the $r={\rm const}.$ hypersurfaces
\begin{align}
      ds^2 = -e^{2\nu(\chi)}dt^2 + d\chi^2 + R^2(\chi)d\Omega^2,
      \label{an1}
\end{align}
where $t\geq 0,\; \chi\geq 0,$ and $d\Omega^2$ is the standard metric on the unit sphere.
The resulting field equations $G_{ab} + \Lambda g_{ab} = 8\pi T_{ab}$
are given by
\begin{align}
      \frac{1-R'^2-2RR''}{R^2}-\Lambda &= 8\pi \rho,
      \label{an2} \\
      \frac{R'^2-1+2RR'\nu'}{R^2} + \Lambda &= 8\pi p,
      \label{an3} \\
      \frac{\nu' R' +R(\nu'^2 +\nu'')+R''}{R} +\Lambda &= 8\pi p_{\perp}.
      \label{an4}
\end{align}
Here $\rho,\, p$ and $p_{\perp}$ are the energy density, the radial pressure and the tangential pressure respectively. In the present paper we assume that $\rho\geq 0,\;p\geq 0,$ and that the energy condition
\begin{equation}\label{encond2}
p+2p_{\perp}\leq \rho,
\end{equation}
holds.
We are only interested in solutions with a regular centre and we therefore impose the conditions
\begin{equation}\label{regularcentre}
R(0)=0,\;\; R'(0)=1.
\end{equation}
The invariant mass function in spherically symmetric cosmological spacetimes can be defined as
\begin{align}
      m(\chi)=\frac{R(\chi)}{2}\left[1-R'^2(\chi)\right]
      - \frac{\Lambda}{6}R^3(\chi),
      \label{anmass}
\end{align}
where the prime denotes differentiation with respect to $\chi$. Note that (\ref{regularcentre}) implies
that $$\lim_{\chi\to 0}\frac{m(\chi)}{R(\chi)}=0.$$
Since we consider non-isotropic solutions the Tolman-Oppenheimer-Volkov equation needs to be modified and becomes
\begin{align}\label{an5}
  p' = -(\rho+p)(4\pi p + \frac{m}{R^3}-\Lambda/3)\frac{R}{R'}
  -2\frac{R'}{R}(p-p_{\perp}).
\end{align}
We can now state our main result.
\begin{theorem}\label{theorem1}
Let $\Lambda\geq 0$ be given and assume that a solution of the Einstein equations (\ref{an2})--(\ref{an4})
exists on an interval $[0,\chi_b],$ and satisfies (\ref{encond2}) and (\ref{regularcentre}). Given $\chi$ with $0<\chi\leq \chi_b,$ then if
\begin{equation}\label{lambdassumption}
\Lambda R^2(\chi)\leq 1,
\end{equation}
it holds that
\begin{equation}\label{thmineq}
\frac{m(\chi)}{R(\chi)}\leq\frac29-\frac{\Lambda R(\chi)^2}{3}+\frac29 \sqrt{1+3\Lambda R(\chi)^2}.
\end{equation}
\end{theorem}
Note in particular that if $R=R(\chi_b)$ denotes the boundary of the static object then the inequality
takes the form as in (\ref{ineqlambda}).
Let us now check that the inequality (\ref{thmineq}) is consistent with equation (\ref{anmass}) 
which can be written as
\begin{align}\label{Rprime}
(R')^2=1-\frac{2m}{R}-\frac{\Lambda R^2}{3}.
\end{align}
It is clear that if too large values of $2m/R$ were allowed then this equation would not be meaningful.
However, the inequality (\ref{thmineq}) guarantees that the right hand side of (\ref{Rprime}) is always positive.
Indeed, we have the following result.
\begin{corollary}\label{corollary1}
The inequality (\ref{thmineq}) implies that
\begin{align}
1-\frac{2m}{R}-\frac{\Lambda R^2}{3}>0,
\end{align}
when $\Lambda R^2<1.$
\end{corollary}
\textit{Proof of Corollary~\ref{corollary1}: }
We have from (\ref{thmineq}) that
\begin{align}
\frac{2m}{R}+\frac{\Lambda R^2}{3}\leq \frac49-\frac{\Lambda R^2}{3}+\frac49\sqrt{1+3\Lambda R^2}.
\end{align}
It is easy to see that the right hand side is an increasing function of $\Lambda R^2$ in the interval $[0,1],$ and the right hand side equals $1$ when $\Lambda R^2=1.$ This completes the proof of the corollary.
\begin{flushright}
$\Box$
\end{flushright}
Next we turn to the issue of sharpness. As was mentioned in the introduction, infinitely thin shell solutions saturate the Buchdahl inequality, i.e. the case $\Lambda=0,$ as well as the inequality derived in \cite{An5} for charged static objects. Thus it is natural to investigate if infinitely thin shell solutions also saturate (\ref{thmineq}). As we will see below this is generally not the case.
Let us for this purpose consider a sequence of regular shell solutions which approach an infinitely thin shell. More precisely, by a regular solution $\Psi=(p,p_{\perp},\rho,\nu,R)$ of the Einstein equations we mean that $R$ and $\nu$ are $C^2$ except at finitely many points, that the matter quantities $p,p_{\perp}$ and $\rho$ are $C^1$ except at finitely many points, $p$ has compact support and the equations (\ref{an2})--(\ref{an4}) and (\ref{an5}) are satisfied almost everywhere. Now let $\Psi_k:=(p_k,(p_{\perp})_k,\rho_k,\nu_k,R_k)$ be a sequence of regular solutions such that the matter terms
$p_k,(p_{\perp})_k$ and $\rho_k$ have support in $[\chi_0^k,\chi_1],$ where
\begin{equation}
\lim_{k\to\infty}\frac{\chi_0^k}{\chi_1}=1.\label{hypothesis}
\end{equation}
Assume that
\begin{align}\label{supofp}
\|R_k^2p^k\|_{\infty}<C, \mbox{ where }C\mbox{ is independent on }k,
\end{align}
and
\begin{align}\label{rhominuspperp}
\int_{\chi_0^k}^{\chi_1}(\rho-2p_{\perp})R^2d\chi\to 0,\mbox{ as }k\to \infty.
\end{align}
Furthermore, denote by $M_k$ the total ADM mass of the solution and assume that $M=\lim_{k\to\infty}M_k$ exists, and assume that
\begin{align}\label{seq}
R_k(\chi_0^k)\to R_1\mbox { as }k\to\infty, \mbox{ where }R_1:=R_k(\chi_1) \mbox{ for all }k.
\end{align}
We can now state our second result.
\begin{proposition}\label{proposition}
Assume that $\{\Psi_k\}_{k=1}^{\infty}$ is a sequence of regular
solutions with the properties specified above.
Then
\begin{equation}
\frac{M}{R_1}=\frac29-\frac{\Lambda R_1^2}{3}+\frac29 \sqrt{1+3\Lambda R_1^2}-H(\Lambda,R_1,M),
\end{equation}
where $H>0$ when $0<\Lambda R_1^2<1,$ and $H=0$ if $\Lambda R_1^2=1$ or $\Lambda R_1=0.$
\end{proposition}
\textbf{Remark 1: }
It is thus clear that an infinitely thin shell with $0<\Lambda R_1^2<1$ will not saturate the inequality. The two cases which give sharpness in the inequality, i.e. when $H=0,$ belong to the boundary of our domain and these should be treated as limits of sequences. For instance, in the case $R_1=0$ we think of a sequence $\{R_1^j\}_{j=1}^{\infty},$ such that $R_1^j\to 0$ as $j\to\infty,$ and for each fixed $j$ we consider a sequence of thin shells which approach an infinitely thin shell at $R=R_1^j.$ Likewise for the case $\Lambda R_1^2=1.$ In the former case the influence of $\Lambda$ becomes negligible since when $R_1\to 0,$ $\Lambda R^2_1\to 0,$ and an infinitely thin shell at $R_1=0$ will clearly saturate the inequality since it reduces to the Buchdahl case, cf. \cite{An1}. In the latter situation we have $M/R_1=1/3,$ which is considerably lower than the maximum value $4/9$ when $\Lambda=0.$ However, in contrast to the case with vanishing cosmological constant where the limit $M/R_1=4/9$ implies that the red shift factor is bounded by $2,$ the case when $\Lambda R_1^2$ approaches $1$ does not provide a bound and the red shift factor can be arbitrarily large. Recall here that a bound on the red shift follows from a bound on
$$\frac{1}{\sqrt{1-\frac{2M}{R_1}-\frac{\Lambda R^2_1}{3}}}.$$
\textbf{Remark 2: }
That sequences exist with the properties specified in the proposition has been proved for
the Einstein-Vlasov system in the case $\Lambda=0,$ cf. \cite{An2} and \cite{AR2} for a numerical study. It is interesting to note that the sequence of shells constructed in \cite{An2}, which approach an infinitely thin shell, have support in $[R_0^j,R_0^j(1+(R_0^j)^q)],\, q>0,$ where $R_0^j\to 0$ as $j\to\infty.$ Hence, this sequence gives in the limit an infinitely thin shell with $R_1=0,$ which corresponds to the degenarate case discussed in Remark 1 above.\\

\section{Proof of Theorem \ref{theorem1}}

As mentioned in the introduction our method of proof is an adaption of the method in \cite{KS} to the case with a positive cosmological constant. Let us introduce the following variables
\begin{align}
  x &= \frac{2m}{R} + \frac{\Lambda}{3}R^2 = 1 - R'^2,\\
  y &= 8\pi R^2 p,\\
  z &= \Lambda R^2.
\end{align}
Furthermore, we introduce a new independent variable
\begin{align}
  \beta = 2 \log R(\chi),
\end{align}
and we denote the derivative with respect to $\beta$ by a dot. Note that this is a valid transformation of variables. Indeed, $R'(0)=1$ implies that $R$ is an increasing function of $\chi$ in an interval
$0\leq\chi <\epsilon.$ Assume that this is the maximal interval on which $R'>0,$ and assume that $\Lambda \epsilon^2<1,$ and that $\epsilon<\chi_b.$ The arguments given below then lead to the conclusion of Theorem \ref{theorem1} for $\chi\leq\epsilon.$ In view of Corollary \ref{corollary1} and (\ref{Rprime}) we then have
\begin{align}\label{Rprime2}
R'(\chi)=\sqrt{1-\frac{2m(\chi)}{R(\chi)}-\frac{\Lambda R(\chi)^2}{3}}>0,
\end{align}
on $0\leq\chi\leq \epsilon.$ Thus $R'(\epsilon)>0$ which shows that either $\epsilon\geq\chi_b$ or $\Lambda\epsilon^2\geq 1.$ Hence, $R$ is an increasing function of $\chi$ in the admissible domain.
The Einstein field equations (\ref{an2})--(\ref{an4}) can now be rewritten as follows
\begin{align}
  2\dot{x} + x - z &= 8\pi \rho R^2,\label{ode1}\\
  y &= 8\pi p R^2,\label{ode2}\\
  \frac{\dot{x}}{2(1-x)}(x+y-z) + \dot{y} + \frac{(x+y-z)^2}{4(1-x)}
  &= 8\pi p_{\perp} R^2.\label{ode3}
\end{align}
It should also be noted that
\begin{align}
  \dot{z} = \frac{dz}{d\beta} = \frac{1}{2}\frac{R}{R'}\frac{dz}{d\chi} =
  \frac{1}{2}\frac{R}{R'}(\Lambda 2 R R') = \Lambda R^2 = z.\label{ode4}
\end{align}

Expressing $\rho, p$ and $p_{\perp}$ by the equations (\ref{ode1})--(\ref{ode3}), the energy condition
\begin{align}
  p + 2p_{\perp}\leq \rho
\end{align}
becomes
\begin{align}
  y + \frac{\dot{x}}{(1-x)}(x+y-z) + 2\dot{y} + \frac{(x+y-z)^2}{2(1-x)}
  \leq 2\dot{x} + x - z.
\end{align}

Reordering of the terms and using (\ref{ode4}) we obtain the inequality
\begin{eqnarray}
  &(3x+y-2-z)\dot{x} + 2(1-x)\dot{y} -2(1-x)\dot{z} &\nonumber \\
  &\leq -\frac{1}{2}\left[3x^2 + (y-z)^2 - 2(x-y) + 2z(3-4x)\right] :=
  -\frac{1}{2}u(x,y,z).&
\end{eqnarray}

Next, let us define
\begin{align}
  w = \frac{(3(1-x)+1+y-z)^2}{(1-x)},
\end{align}
from which we can compute
\begin{align}\label{ineq}
  \dot{w} &= \frac{4-3x+y-z}{(1-x)^2}\left[(3x+y-2-z)\dot{x} +
  2(1-x)\dot{y} -2(1-x)\dot{z}\right]\\
  &\leq -\frac{4-3x+y-z}{2(1-x)^2} u(x,y,z).\label{wdot}
\end{align}
Note that $0 \leq x \leq 1$ and $y > 0$, and from the restriction (\ref{lambdassumption}) on $\Lambda$ we also have that $0\leq z\leq 1.$ This latter condition is important to fix the sign of the factor in front of $u$ in (\ref{wdot}) to ensure the validity of the optimization problem below.

In view of (\ref{ineq}) we thus find that $w$ is decreasing if $u$ is positive and hence
\begin{align}
  w \leq \max_{0 \leq x \leq 1, y\geq 0, 0 \leq z \leq 1, u \leq 0} w(x,y,z).\label{opt1}
\end{align}
We now show that the solution of this optimization problem is $w=16$ attained at $x=y=z=0.$ First note that $u=0$ at the centre of symmetry, since $x=y=z=0$ there, so that our domain is nonempty. The condition
$u\leq 0$ can be written as
\[
3x(x-1)+x-8zx+(y-z)^2+2y+6z\leq 0,
\]
which is equivalent to
\begin{eqnarray}
& &(3x-8z+1)(x-1)-8z+1+(y-z)^2+2y+6z\nonumber\\
& &=(3x-8z+1)(x-1)+(y-z+1)^2\leq 0.
\end{eqnarray}
Thus we have
\begin{equation}\label{uconstraint}
(1+y-z)^2\leq (3x-8z+1)(1-x).
\end{equation}
Hence
\begin{eqnarray}
w&=&9(1-x)+6(1+y-z)+\frac{(1+y-z)^2}{1-x}\nonumber\\
&\leq& 9(1-x)+6(1+y-z)+(3x-8z+1)=16-6x+6y-14z.\nonumber
\end{eqnarray}
From (\ref{uconstraint}) we also have
\[
2(y-z)\leq (3x-8z+1)(1-x)-1-(y-z)^2\leq (3x-8z+1)(1-x)-1,
\]
and we obtain
\begin{eqnarray}\label{wmax}
w&\leq& 16-6x+6(y-z)-8z=16-6x+3(3x-8z+1)(1-x)-3-8z\nonumber\\
&=&16-9x^2-24z(1-x)-8z\leq 16.
\end{eqnarray}
The point $(0,0,0)$ is admissible since $u(0,0,0)=0,$ and moreover,
$w(0,0,0)=16,$ which proves the claim above.
We thus get
\begin{align}\label{xzineq}
  (3(1-x)+1-z)^2 \leq 16 (1-x).
\end{align}
We introduce the dimensionless variables
\begin{align}
  X = \frac{m(\chi)}{R(\chi)},
\end{align}
so that the inequality reads
\begin{align}\label{xz}
  \left(\frac{3}{2}X + \frac{1}{2}z\right)^2 \leq
  \frac{2}{3}\left(\frac{3}{2}X + z\right).
\end{align}
This can be written as
\begin{align}
\left(X-\frac29+\frac{z}{3}-\frac23\sqrt{\frac19+\frac{z}{3}}\right)\left(X-\frac29+\frac{z}{3}+\frac23\sqrt{\frac19+\frac{z}{3}}\right)\leq 0.
\end{align}
The second factor is non-negative and vanishes only when $X=z=0$ which implies that
\begin{equation}
X\leq \frac29-\frac{z}{3}+\frac23\sqrt{\frac19+\frac{z}{3}}.\label{ineq1}
\end{equation}
By inserting the expressions for $X$ and $z$ one obtains
\begin{equation}
\frac{m(\chi)}{R(\chi)}\leq \frac29-\frac{\Lambda R^2(\chi)}{3}+\frac29\sqrt{1+3\Lambda R^2(\chi)},
\end{equation}
which is the claimed inequality. This completes the proof of Theorem \ref{theorem1}.
\begin{flushright}
$\Box$
\end{flushright}
Since the cosmological constant is regarded to be a small quantity, in the sense that $3\Lambda R^2<<1,$ it is interesting to make a Taylor expansion of the right hand side which implies that
\begin{equation}\label{Taylorexp}
\frac{m}{R}\leq\frac49-\frac{\Lambda^2 R^4}{4}.
\end{equation}
Hence, the influence of $\Lambda$ is of the second order.

\section{Proof of Proposition \ref{proposition}}

The proof uses the ideas in \cite{An3}. However, the arguments are slightly different, in particular due to lack of monotonicity of $\nu$ when $\Lambda>0.$
We define
\begin{align}
  \Gamma_k := (4\pi p_k R_k^3(\chi) + m_k - \frac{\Lambda}{3}R_k^3(\chi))\frac{e^{\nu_k}}{R_k'(\chi)}
  \label{eqn:ga1}.
\end{align}
We then have
\begin{align}
  \Gamma'_k = (4\pi (\rho_k + p_k + 2(p_{\perp})_k)R_k^2(\chi) - \Lambda R_k^2(\chi)) e^{\nu_k}.
  \label{eqn:ga2}
\end{align}
Below we sometimes drop the index $k$ but it is inserted when we find it necessary for clarity.

From the first field equation we find
\begin{align}
  R'' = \partial_\chi R' = -4\pi\rho R + \frac{m}{R^2} - \frac{\Lambda}{3}R.
  \label{eqn:rpp}
\end{align}

Let us integrate Eq.~(\ref{eqn:ga1}) with respect to $\chi$ in the interval $[\chi_0,\chi_1].$ This leads to
\begin{eqnarray}
  \Gamma(\chi_1)-\Gamma(\chi_0) &=& \int_{\chi_0}^{\chi_1}\left[
  4\pi (\rho + p + 2p_{\perp})R^2 - \Lambda R^2
  \right] e^\nu d\chi\nonumber \\
  &=&e^{\nu(\xi)}\int_{\chi_0}^{\chi_1}\left[
  4\pi (\rho + p + 2p_{\perp})R^2 - \Lambda R^2
  \right] d\chi,
\end{eqnarray}
where $\xi\in [\chi_0,\chi_1].$
By using the energy condition
\begin{align}
  p + 2p_{\perp}\leq\rho,
\end{align}
together with the condition (\ref{rhominuspperp}) we obtain
\begin{align}
  \Gamma(\chi_1)-\Gamma(\chi_0) &=e^{\nu(\xi)} \int_{\chi_0}^{\chi_1}\left[
  4\pi (2\,\rho)R^2 - \Lambda R^2\right] d\chi + o(k^{-1})\\
  & = e^{\nu(\xi)}\int_{\chi_0}^{\chi_1}\left[-2 R \partial_{\chi} R' + \frac{2m}{R} -\frac53
  \Lambda R^2 \right] d\chi + o(k^{-1}),
\end{align}
where equation~(\ref{eqn:rpp}) was taken into account. Here $o(k^{-1})$ is used for terms which vanish in the limit $k \to \infty.$
Since we are interested in the limit $k\to\infty,$ and since $\chi_0^k\to\chi_1,$ as $k\to\infty,$
we note that
\begin{align}
\int_{\chi_0^k}^{\chi_1}\frac{2m_k}{R_k} -\frac53
  \Lambda R_k^2 d\chi\to 0\mbox{ as }k\to\infty,
\end{align}
and this term will therefore be included in the $o(k^{-1})$ term.
The first integral is easily evaluated and we obtain
\begin{eqnarray}
&e^{\nu(\xi)}\int_{\chi_0}^{\chi_1}\left[-2 R \partial_{\chi} R'\right] d\chi &\nonumber \\
&=
e^{\nu(\xi)}\left(2R(\chi_0)\sqrt{1-\frac{\Lambda R^2(\chi_0)}{3}}-2R_1\sqrt{1-\frac{2M}{R_1}-\frac{\Lambda R^2_1}{3}}\right)\nonumber \\
&+e^{\nu(\xi)}\int_{\chi_0}^{\chi_1}2(R')^2d\chi.
\end{eqnarray}
Now $(R_{k}')^2$ is bounded by $1$ in view of (\ref{Rprime}) and therefore the last integral vanishes in the limit $k\to\infty.$
Let us next show that
\begin{align}\label{nu1}
e^{\nu_k(\xi_k)}\to\sqrt{1-\frac{2M}{R_1}-\frac{\Lambda R^2_1}{3}}.
\end{align}
From the Einstein equations we have
\begin{align}
\nu_{k}'=\frac{4\pi R_k p_k+\frac{m_k}{R_{k}^2}-\frac{\Lambda R_{k}^2}{3}}{R_{k}'},
\end{align}
so that
\begin{align}\label{nucontinuous}
\nu_k(\chi_1)-\nu_k(\xi_k)=\int_{\xi_k}^{\chi_1}\frac{4\pi R_k p_k+\frac{m_k}{R_{k}^2}-\frac{\Lambda R_{k}^2}{3}}{R_{k}'}d\chi.
\end{align}
Now since $\Lambda R_k^2(\chi_1)<1$ it follows from the argument following the formulation of Theorem~\ref{theorem1} that $$R_{k}'(\chi)=\sqrt{1-\frac{2m_k}{R_k}-\frac{\Lambda R_k^2}{3}}>0, \mbox{ for all }\chi\leq\chi_1.$$
Moreover, from the assumption (\ref{supofp}) and the general fact that $m_k/R_k\leq 4/9$ we get
\begin{eqnarray}
& &\int_{\xi_k}^{\chi_1}\frac{4\pi R_k p_k+\frac{m_k}{R_{k}^2}-\frac{\Lambda R_{k}^2}{3}}{R_{k}'}d\chi \leq
C\int_{\xi_k}^{\chi_1}\frac{1}{R_k}d\chi\nonumber \\
& &=C\int_{R_1}^{R_k(\chi_1)}\frac{1}{R_kR_{k}'}dR_k\leq C\int_{R_k(\xi_k)}^{R_1}\frac{1}{R_k}dR_k\nonumber \\
& &=C\log{\frac{R_1}{R_k(\xi_k)}}\to 0\mbox{ as }k\to\infty,
\end{eqnarray}
where we used the assumption (\ref{seq}) for the final conclusion.
The claim (\ref{nu1}) follows since for all $k$ we have
$$e^{\nu_k(\chi_1)}=\sqrt{1-\frac{2M}{R_1}-\frac{\Lambda R^2_1}{3}},$$
since there is no matter in the region $R>R_1.$
This result together with the condition (\ref{seq}), i.e.,
\begin{align}\label{Rcont}
R_k(\chi_0^k)\to R_1, \mbox{ as } k\to \infty,
\end{align}
implies that
$$\Gamma_k(\chi_0^k)=\frac{\Lambda R_1^3\sqrt{1-\frac{2M}{R_1}-\frac{\Lambda R^2_1}{3}}}{3\sqrt{1-\frac{\Lambda R_1^2}{3}}}+o(k^{-1}).$$
Thus to summarize we have obtained
\begin{eqnarray}
\Gamma_k(\chi_1)&=&
\sqrt{1-\frac{2M}{R_1}-\frac{\Lambda R^2_1}{3}}\left(2R_1\sqrt{1-\frac{\Lambda R^2_1}{3}}-2R_1\sqrt{1-\frac{2M}{R_1}-\frac{\Lambda R^2_1}{3}}\right)\nonumber \\
& &-\frac{\Lambda R_1^2\sqrt{1-\frac{2M}{R_1}
-\frac{\Lambda R^3_1}{3}}}{3\sqrt{1-\frac{\Lambda R_1^2}{3}}}+o(k^{-1}).
\end{eqnarray}
Here we again used (\ref{Rcont}).
In view of the condition that $\lim_{k\to\infty}m_k=M$ we obtain in the limit after some rearranging
\begin{align}\label{hequality}
\frac{M}{R_1}-\frac{\Lambda R_1^2}{3}=
2\sqrt{1-\frac{2M}{R_1}-\frac{\Lambda R^2_1}{3}}\left(1-\sqrt{1-\frac{2M}{R_1}-\frac{\Lambda R^2_1}{3}}\right)-h(\Lambda,R_1,M),
\end{align}
where
\begin{align}\label{h}
h(\Lambda,R_1,M):=\frac{\Lambda R_1^2\sqrt{1-\frac{2M}{R_1}-\frac{\Lambda R^2_1}{3}}}{\sqrt{1-\frac{\Lambda R^2_1}{3}}}.
\end{align}
It should now be noted that if $h=0,$ the expression (\ref{hequality}) is equivalent to
\begin{align}
\left(\frac{3}{2}\frac{M}{R_1} + \frac{1}{2}\Lambda R_1^2\right)^2=
  \frac{2}{3}\left(\frac{3}{2}\frac{M}{R_1} + \Lambda R_1^2\right),
\end{align}
which in view of (\ref{xz}) leads to equality in (\ref{thmineq}). If we carry out the algebra we find that (\ref{hequality}) leads to the following inequality
\begin{align}\label{ineqH}
\frac{M}{R_1}\leq\frac29-\Lambda R_1^2+\frac29 \sqrt{1+3\Lambda R_1^2}-H(\Lambda,R_1,M),
\end{align}
where
\begin{align}\label{H}
H=\frac{(2\sqrt{1-\frac{2M}{R_1}-\frac{\Lambda R_1^2}{3}}-h)h}{\frac{M}{R_1}-\frac29+\frac{\Lambda R_1^2}{3}+\frac29\sqrt{1+3\Lambda R_1^2}}.
\end{align}
It is straightforward to check that $H>0$ when $0<\Lambda R^2_1<1,$ and that $H=0$ when $R_1=0$ or $1-2M/R_1-\Lambda R_1^2/3=0.$ In the latter case, since $H=0$ we have equality in (\ref{thmineq}), and from the proof of Corollary~\ref{corollary1} we thus find that necessarily $\Lambda R_1^2=1.$ Thus $H=0$ when $\Lambda R_1^2=1.$ This completes the proof of Proposition~\ref{proposition}.
\begin{flushright}
$\Box$
\end{flushright}



\end{document}